\documentclass[12pt]{article}
 \usepackage{hyperref,amsfonts,amssymb,amsmath,mathrsfs}
 \usepackage[english]{babel}

\def\noi{\noindent}
\def\nqq{\hspace{-2em}}

\def\barr{\left(\begin{array}}
\def\earr{\end{array}\right)}
\def\beq#1{\begin{equation}\label{#1}}
\def\eeq{\end{equation}}
\def\ber#1{\begin{eqnarray}\label{#1} &&\nqq}
\def\eer{\end{eqnarray}}

\newcommand{\bear}[1]{\begin{eqnarray}\label{#1}}
\newcommand{\ear}{\end{eqnarray}}

\catcode`\@=11 \@addtoreset{equation}{section}\catcode`\@=12

\newcommand{\SL}{\mathop{\rm sl}\nolimits}
\newcommand{\SO}{\mathop{\rm so}\nolimits}
\newcommand{\SP}{\mathop{\rm sp}\nolimits}

\newcommand{\fnm}{\footnotemark}
\newcommand{\fnt}{\footnotetext}

\begin{document}

 \vspace{15pt}

 \begin{center}
 \large\bf

 On calculation of  fluxbrane polynomials
  corresponding to classical series of Lie algebras

 \vspace{15pt}

 \normalsize\bf
              A. A. Golubtsova\fnm[1]\fnt[1]{siedhe@gmail.com}$^{, b}$
           and   V. D. Ivashchuk\fnm[2]\fnt[1]{rusgs@phys.msu.ru}$^{, a, b}$

 \vspace{7pt}

 \it (a) \ \ \ Center for Gravitation and Fundamental
 Metrology,  VNIIMS, 46 Ozyornaya Str., Moscow 119361, Russia  \\

 (b) \  Institute of Gravitation and Cosmology,
 Peoples' Friendship University of Russia,
 6 Miklukho-Maklaya Str.,  Moscow 117198, Russia \\

 \end{center}
 \vspace{15pt}

 \small\noi

 \begin{abstract}
  We present a description of computational
  program (written in Maple) for calculation of fluxbrane polynomials
  corresponding to classical simple Lie algebras. These polynomials
  define certain special solutions to open Toda chain equations.

 \end{abstract}

 \section{Introduction}

  In this paper we deal with a set of equations
  \beq{1.1}
  \frac{d}{dz} \left( \frac{ z}{H_s} \frac{d}{dz} H_s \right) =
   P_s \prod_{s' = 1}^{r}  H_{s'}^{- A_{s s'}},
  \eeq
 with  the following boundary conditions imposed:
 \beq{1.2}
   H_{s}(+ 0) = 1,
 \eeq
 $s = 1,...,r$. Here the functions $H_s(z) > 0$ are
 defined on the interval $(0, +\infty)$,
 $P_s > 0$  for all $s$ and $(A_{s s'})$ is the Cartan matrix
 for some finite dimensional  simple Lie algebra $\cal G$
 of rank $r$ ($A_{ss} = 2$ for all $s$).

 The functions $H_s(z) > 0$ appear as moduli functions of
 generalized fluxbrane solutions obtained in  \cite{Iflux}.
 Parameters  $P_s$ are proportional
 to brane charge density squared $Q_s^2$ and $z = \rho^2$, where
  $\rho$ is a radial parameter. The boundary condition
 (\ref{1.2}) guarantees the absence of singularity
 (in the metric) for $\rho =  +0$.  For fluxbrane solutions see \cite{Iflux},
  \cite{GW}-\cite{GIM} and references therein. (The more general
  classes of solutions were described in \cite{IK,IMtop}). The simplest
 ``fluxbrane'' solution  is a well-known Melvin solution
  \cite{Melv} describing the gravitational field of a flux
  tube. The Melvin solution corresponds to the Lie algebra $A_1 = sl(2)$
  of rank $1$.

 It was conjectured in \cite{Iflux} that eqs. (\ref{1.1}),
 (\ref{1.2})  have polynomial solutions
 \beq{1.3}
  H_{s}(z) = 1 + \sum_{k = 1}^{n_s} P_s^{(k)} z^k,
 \eeq
 where $P_s^{(k)}$ are constants, $k = 1,\ldots, n_s$. Here
 $P_s^{(n_s)} \neq 0$  and
 \beq{1.4}
  n_s = 2 \sum_{s' =1}^{r} A^{s s'}.
 \eeq
 $s = 1,...,r$,  where  $(A^{s s'}) = (A_{s s'})^{-1}$.
 Integers $n_s$ are components  of the so-called twice dual
 Weyl vector in the basis of simple roots \cite{FS}.
 It was pointed in \cite{Iflux} that the conjecture on polynomial
 structure of $H_s$  (suggested originally for semisimple Lie algebras) may be
 verified for $A_n$ and $C_n$ Lie algebras along a line as it
 was done for black-brane polynomials from \cite{IMbb} (see also \cite{IMtop}).
In \cite{Iflux} certain examples of fluxbrane solutions
corresponding  to Lie algebras $A_1 \oplus \ldots \oplus A_1$ and
$A_2$ were presented.

The substitution of (\ref{1.3})  into (\ref{1.1}) gives an
infinite chain of relations on parameters $P_s^{(k)}$  and
 $P_s$.  The first relation in this chain
 \beq{1.5a}
  P_s  =  P_s^{(1)},
 \eeq
 $s = 1, ..., r$, corresponds to $z^0$-term in the decomposition
 of (\ref{1.1}).

 {\bf Special solutions.} We note that for a special choice
 of $P_s$ parameters: $P_s = n_s P$, $P > 0$, the polynomials
 have the following simple form \cite{Iflux}
 \beq{1.s}
 H_{s}(z) = (1 +  P z)^{n_s},
 \eeq
 $s = 1, ..., r$. This relation may be considered as nice
 tool for verification of general solutions obtained
 by either  analytical or computer calculations.

 {\bf Remark: open Toda chains.} It should be also noted that
  a set of polynomials $H_s$ define a special solution to the open Toda chain
  equations \cite{T,B,K,OP} corresponding to the Lie algebra $\cal G$
 \beq{1.T}
 \frac{d^2 q^s}{du^2} = -  B_s \exp( \sum_{s' = 1}^{r} A_{s s'} q^{s'} ),
 \eeq
  where $B_s = 4 P_s$,
 \beq{1.h}
  H_s = \exp(- q^s(u) -  n_s u),
 \eeq
  $s = 1, ..., r$ and $z = e^{-2u}$.

 In this paper we suggest a computational program for calculations
 of polynomials  corresponding  to classical series of simple Lie algebras.

\section{Cartan matrices for classical simple Lie algebras}

 Here we list, for convenience, the Cartan matrices for all classical simple Lie algebras
 and inverse Cartan matrices as well.

 In summary \cite{FS}, there are four classical infinite
 series of simple Lie algebras, which are denoted by

 \ber{Ap.9}
  A_r \ (r \ge 1), \quad B_r \ (r \ge 3), \quad C_r \ (r \ge 2), \quad D_r \
  (r \ge 4).
 \eer

  In all cases the subscript denotes the rank of the algebra.
 The algebras in the infinite series of simple Lie algebras are
 called the classical (Lie) algebras. They are isomorphic to the matrix algebras

 \ber{Ap.11}
  A_r \cong\SL(r+1), \quad B_r\cong\SO(2r+1), \quad
  C_r \cong\SP(r), \quad D_r\cong\SO(2r).
  \eer

  {\bf $A_r$ series.} Let $A = \left(A_{ss'}\right)$ be $r \times r$
 Cartan matrix for the Lie algebra $A_r= \SL(r+1)$, $r \ge 1$.
  The Cartan matrices for $A_r$-series have the following form
 \beq{A.1a}
 \left(A_{ss'}\right)= \left( \begin{array}{*{6}{c}}
 2&-1&0&\ldots&0&0\\
 -1&2&-1&\ldots&0&0\\
 0&-1&2&\ldots&0&0\\
 \multicolumn{6}{c}{\dotfill}\\
 0&0&0&\ldots&2&-1\\
 0&0&0&\ldots&-1&2
 \end{array}
 \right)\quad
 \eeq

This matrix is described graphically by the Dynkin diagram
pictured on Fig. 1.
\begin{center}
\bigskip
\begin{picture}(66,11)
\put(5,10){\circle*{2}} \put(15,10){\circle*{2}}
\put(25,10){\circle*{2}} \put(55,10){\circle*{2}}
\put(65,10){\circle*{2}} \put(5,10){\line(1,0){30}}
\put(45,10){\line(1,0){10}} \put(5,5){\makebox(0,0)[cc]{1}}
\put(15,5){\makebox(0,0)[cc]{2}} \put(25,5){\makebox(0,0)[cc]{3}}
\put(40,10){\makebox(0,0)[cc]{$\dots$}}
\put(55,5){\makebox(0,0)[cc]{$r-1$}}
\put(65,5){\makebox(0,0)[cc]{$r$}} \put(55,10){\line(1,0){10}}
\end{picture} \\[5pt]
\small Fig. 1. \it Dynkin diagram for $A_r$ Lie algebra
\end{center}

 For $s \ne s'$, $A_{ss'}=-1$ if nodes $s$ and $s'$ are connected
by a line on the diagram and $A_{ss'}=0$ otherwise. Using the
relation for the inverse matrix $A^{-1}=(A^{ss'})$ (see Sect.7.5
in \cite{FS})
 \ber{Ap.12}
  A^{ss'} = \frac1{r+1}\min(s,s')[r+1-\max(s,s')]
  \eer we may
 rewrite (\ref{1.4}) as follows
   \ber{Ap.13}
   n_s = s(r+1 -s),
   \eer
   $s= 1,\dots,r$.

{\bf $B_r$ and $C_r$ series.} For $B_r$-series we have
 the following Cartan matrices
 \beq{B.1a}
 \left(A_{ss'}\right)= \left( \begin{array}{*{6}{c}}
 2&-1&0&\ldots&0&0\\
 -1&2&-1&\ldots&0&0\\
 0&-1&2&\ldots&0&0\\
 \multicolumn{6}{c}{\dotfill}\\
 0&0&0&\ldots&2&-2\\
 0&0&0&\ldots&-1&2
\end{array}
\right)\quad
 \eeq
 while for $C_r$-series  the Cartan matrices read as follows
 \beq{C.1a}
 \left(A_{ss'}\right)= \left( \begin{array}{*{6}{c}}
 2&-1&0&\ldots&0&0\\
 -1&2&-1&\ldots&0&0\\
 0&-1&2&\ldots&0&0\\
 \multicolumn{6}{c}{\dotfill}\\
 0&0&0&\ldots&2&-1\\
 0&0&0&\ldots&-2&2
\end{array}
\right)\quad
 \eeq

Dynkin diagrams for these cases are pictured on Fig. 2.
\begin{center}
\bigskip
\begin{picture}(66,12)
\put(5,10){\circle*{2}} \put(15,10){\circle*{2}}
\put(25,10){\circle*{2}} \put(55,10){\circle*{2}}
\put(65,10){\circle*{2}} \put(5,10){\line(1,0){30}}
\put(45,10){\line(1,0){10}} \put(5,5){\makebox(0,0)[cc]{1}}
\put(15,5){\makebox(0,0)[cc]{2}} \put(25,5){\makebox(0,0)[cc]{3}}
\put(40,10){\makebox(0,0)[cc]{$\dots$}}
\put(55,5){\makebox(0,0)[cc]{$r-1$}}
\put(65,5){\makebox(0,0)[cc]{$r$}} \put(55,11){\line(1,0){10}}
\put(55,9){\line(1,0){10}} \put(60,10){\makebox(0,0)[cc]{\large
$>$}}
\end{picture} \qquad

\begin{picture}(66,12)
\put(5,10){\circle*{2}} \put(15,10){\circle*{2}}
\put(25,10){\circle*{2}} \put(55,10){\circle*{2}}
\put(65,10){\circle*{2}} \put(5,10){\line(1,0){30}}
\put(45,10){\line(1,0){10}} \put(5,5){\makebox(0,0)[cc]{1}}
\put(15,5){\makebox(0,0)[cc]{2}} \put(25,5){\makebox(0,0)[cc]{3}}
\put(40,10){\makebox(0,0)[cc]{$\dots$}}
\put(55,5){\makebox(0,0)[cc]{$r-1$}}
\put(65,5){\makebox(0,0)[cc]{$r$}} \put(55,11){\line(1,0){10}}
\put(55,9){\line(1,0){10}} \put(60,10){\makebox(0,0)[cc]{\large
$<$}}
\end{picture} \\[5pt]
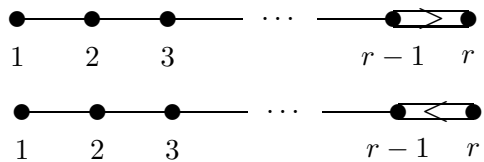
\small Fig. 2. \it Dynkin diagrams for $B_r$ and $C_r$ Lie
algebras, respectively
\end{center}

 In these cases we have the following formulas for inverse
 matricies $A^{-1}=(A^{ss'})$:
 \ber{Ap.14}
  A^{ss'} = \left\{\begin{array}{ll}
  \min(s,s') & \mbox{for } s\ne r, \\[5pt]
  \frac12s' & \mbox{for }s=r,
  \end{array}\right. \quad
  A^{ss'} = \left\{\begin{array}{ll}
  \min(s,s') & \mbox{for }s'\ne r, \\[5pt]
  \frac12s & \mbox{for }s'=r
  \end{array}\right.
  \eer
and relation (\ref{1.4}) takes the form \ber{Ap.15}
 n_s = \left\{\begin{array}{ll}
 s(2r+1 -s) & \mbox{for }s \ne r,\\
 \frac r2(r+1) & \mbox{for }s=r;
 \end{array}\right.\quad
 n_s = s(2r -s), \eer
 for $B_r$ and $C_r$ series,
 respectively, $s = 1,\dots,r$.

{\bf $D_r$ series.} For $D_r$-series  the Cartan matrices read
 \beq{D.1a}
 \left(A_{ss'}\right)= \left( \begin{array}{*{6}{c}}
 2&-1&0&\ldots&0&0\\
 -1&2&-1&\ldots&0&0\\
 0&-1&2&\ldots&0&0\\
 \multicolumn{6}{c}{\dotfill}\\
 0&0&0&\ldots  2&-1&-1\\
 0&0&0&\ldots -1& 2& 0 \\
 0&0&0&\ldots -1& 0& 2
\end{array}
\right)\quad
 \eeq

We have the following Dynkin diagram for this case (Fig. 3):
\begin{center}
\bigskip
\begin{picture}(63,31)
\put(5,20){\circle*{2}} \put(15,20){\circle*{2}}
\put(25,20){\circle*{2}} \put(55,20){\circle*{2}}
\put(5,20){\line(1,0){30}} \put(45,20){\line(1,0){10}}
\put(5,15){\makebox(0,0)[cc]{1}} \put(15,15){\makebox(0,0)[cc]{2}}
\put(25,15){\makebox(0,0)[cc]{3}}
\put(40,20){\makebox(0,0)[cc]{$\dots$}}
\put(55,15){\makebox(0,0)[rc]{$r-2$}} \put(62,27){\circle*{2}}
\put(62,13){\circle*{2}} \put(55,20){\line(1,1){7}}
\put(55,20){\line(1,-1){7}} \put(62,31){\makebox(0,0)[cc]{$r$}}
\put(62,8){\makebox(0,0)[cc]{$r-1$}}
\end{picture} \\[5pt]
\small Fig. 3. \it Dynkin diagram for $D_r$ Lie algebra
\end{center}

and formula for the inverse matrix $A^{-1}=(A^{ss'})$ \cite{FS}:
 \ber{Ap.16} A^{ss'}=\left\{\begin{array}{ll}
 \min(s,s') & \mbox{for }s,s'\notin\{r,r-1\}, \\[5pt]
 \frac12s & \mbox{for }s\notin\{r,r-1\}, \ s'\in\{r,r-1\}, \\[5pt]
 \frac12s' & \mbox{for }s\in\{r,r-1\}, \ s'\notin\{r,r-1\}, \\[5pt]
 \frac14r & \mbox{for }s=s'=r \mbox{ or } s=s'=r-1, \\[5pt]
 \frac14(r-2) & \mbox{for }s=r, \ s'=r-1 \mbox{ or vice versa.}
 \end{array}\right.
 \eer
 The relation (\ref{1.4}) in this case reads \ber{3.17}
 n_s = \left\{\begin{array}{ll}
 s(2r -1 -s) & \mbox{for }s\notin\{r,r-1\},\\
 \frac r2(r-1) & \mbox{for }s\in\{r,r-1\},
\end{array}\right.
\eer $s = 1,\dots,r$.

 For the simple Lie algebras of type $A_r$, $D_r$,
 all roots have the same length, and any two nodes of Dynkin
 diagram are connected by at most one line. In the other cases
 there are roots of two different lengths, the length of the long
 roots being $\sqrt2$ times the length of the short roots for
 $B_r$, $C_r$,  respectively.

\section{Computing of fluxbrane polynomials}

 At the moment the problem of finding  all coefficients
 $P_{s}^{(k)}$ in polynomials $H_s$ analytically is looking
 as  too complicated one.  Thereby it is essential to create a program,
 calculating all the coefficients required.

The algorithm of finding coefficients $P_{s}^{(k)}$  is the
following one:
\begin{itemize}

 \item
 we substitute the polynomials $H_{s}$ into the set of
 differential equations (\ref{1.1}) and reduce the differential equations to a
 set of algebraic equations for $P_{s}^{(k)}$ (expanding the
 equations into degrees of the variable $z$);

 \item  the derived system of algebraic equations is solved in terms of the
 first coefficients $P_{s}^{(1)} = P_s$.

\end{itemize}

We choose the symbolic computational system Maple v.11.01. for
implementation of the discussed algorithm. The standard Maple
packages \verb"LinearAlgebra" and \verb"PolynomialTools" are
used for working with matrices and polynomials appropriately. So
let us start in the following way:

\begin{verbatim}
> with(LinearAlgebra):
> with(PolynomialTools):
\end{verbatim}
At the next step we need to specify the dimension of the Cartan
matrix:

\begin{verbatim}>S:=3:\end{verbatim}

Thus, the dimension of the Cartan matrix is 3. This variable ($S$)
also determines the twice dual Weyl vector dimension and the
number of the differential equations. A variable for the Lie
algebra is entered similarly.
\begin{verbatim}>algn:=bn:\end{verbatim}

Consequently, by default the program calculates the fluxbrane
polynomials for the Lie algebra $B_{n}$ with the Cartan matrix,
the size of which is 3 . We use the standard Maple function
\verb"Matrix" to declare the Cartan matrix.
\begin{verbatim}>A:=Matrix(S,S):\end{verbatim}

It is more convenient to fill the Cartan matrix with the help of
the separate procedure, which constructs matrix elements depending
on the Lie algebras. Let us consider the procedure. There are
three callable variables in it: a Lie algebra, the matrix size and
the Cartan matrix itself.
\begin{verbatim}>AlgLie:=proc(algn,S,CartA:=Matrix(S,S))\end{verbatim}

The matrix elements are constructed  in compliance with Dynkin diagrams. Here we
consider the construction of the matrix elements for the
simple Lie algebras $A_{n}$, $B_{n}$, $C_{n}$ and $D_{n}$. The
following local variables are essential
\begin{verbatim}
local i,mu,nu;
        i := 0; mu := 0; nu:=0;
        mu := S-1; nu:=S-2;
\end{verbatim}
where \verb"i" is an iteration variable and \verb"mu",\verb"nu"
are variables for subscripting in $B_{n}$, $C_{n}$ and $D_{n}$
algebras. By default all elements of the matrix are zeros. Thereby
it is necessary to fill only the elements different from null
according to the Dynkin diagrams. Initially we fill in the
main diagonal of the matrix, because it is identical to the Cartan
matrix of any algebra.
\begin{verbatim} for i to S do CartA[i, i] := 2 end do;\end{verbatim}

Then, using the conditional operator, we consequently
take under consideration the condition on the matrix size
and the fact of belonging to the Lie algebra. We begin with the $A_{n}$ algebra.
 \begin{verbatim}
 if  (S>=1) and algn=an then end if;
 \end{verbatim}

The elements of the secondary diagonals for $A_{n}$ algebra equal
$-1$ according to the Dynkin diagram. Firstly, we fill in the
upper secondary diagonal, and then it is mirrored one. These
actions are performed down inside the previous operator.
\begin{verbatim}
for i to S-1 do CartA[i, i+1] := -1 end do;
for i to S-1 do CartA[i+1, i] := CartA[i, i+1] end do;
\end{verbatim}

The similar actions are run with the conditional operator for the
$B_{n}$, in agreement with the Dynkin diagram for this algebra.
\begin{verbatim}
  if  (S>=3) and algn=bn then
    for i from 1 to (S - 1 ) do CartA[i+1,i]:=-1 end do;
    for i  from 1 to (S - 2) do CartA[i,i+1]:=CartA[i+1,i] end do;
  end if;
\end{verbatim}
But under the Dynkin diagrams for the Lie algebras $B_{n}$ and
$C_{n}$, certain elements are different from $-1$ for the
secondary diagonals. So, let us supplement the preceding lines
with the following matrix element
\begin{verbatim}
 CartA[mu,S]:=-2;
\end{verbatim}
 In  much the same way we have for the Lie algebra $C_{n}$
\begin{verbatim}
  if  (S>=2) and algn=cn then
    for i to S-1 do CartA[i, i+1] := -1 end do;
    for i to S-2 do CartA[i+1, i] := CartA[i, i+1] end do;
    CartA[S, mu] := -2;
  end if;
\end{verbatim}

For the Lie algebra $D_{n}$ certain non-diagonal elements of the
Cartan matrix differ from $zero$, while some elements in the
secondary diagonals are equal to $zero$. According to the Dynkin
diagram for this algebra the elements of the Cartan matrix are
defined in the following way
\begin{verbatim}
if  (S>=4) and algn=dn then
    for i to S-2 do CartA[i, i+1] := -1 end do;
    for i to S-2 do CartA[i+1, i] := CartA[i, i+1] end do;
    CartA[S, nu] := CartA[mu, nu];
    CartA[nu, S] := CartA[S, nu];
end if;
\end{verbatim}

We return the variable  \verb"CartA" and do not forget to set
closing \verb"end" at the end of the procedure. Thus, the Cartan
matrix is filled, depending on the chosen algebra, by a call of
this procedure  with the appropriate parameters.
\begin{verbatim}
>AlgLie(algn, S, A);
\end{verbatim}
 \[
A :=  \left[ {\begin{array}{rrr}
2 & -1 &0 \\
-1 & 2 & -2\\
0 & -1 & 2 \\
\end{array}}
 \right]
\]

Further we declare the twice dual Weyl vector by means of the
standard Maple procedure \verb"Vector".
\begin{verbatim}
>n := Vector[row](1 .. S):
\end{verbatim}

The elements of the inverse Cartan matrix are necessary for
calculating the twice dual Weyl vector's components. The matrix
inversion is done by means of the standard Maple procedure
\verb"MatrixInverse".
\begin{verbatim}
>A1 := MatrixInverse(A);
\end{verbatim}
\[
A1 :=  \left[ {\begin{array}{ccc} {\displaystyle 1} &
{\displaystyle 1} & {\displaystyle 1} \\ [2ex] {\displaystyle 1} &
{\displaystyle 2} & {\displaystyle 2} \\ [2ex]
{\displaystyle \frac12}& {\displaystyle 1} & {\displaystyle \frac32} \\
\end{array}}
 \right]
\]

We use the standard Maple procedure \verb"add" for calculating the
twice dual Weyl vector's components.
\begin{verbatim}
 >for i to S do n[i] := 2*add(A1[i, j], j = 1 .. S) end do:
\end{verbatim}

The coefficients $P_{s}^{(k)}$ are represented by a matrix.
The number of rows of this matrix is the number of the
differential equations (that is  $S$), and the number of columns
is the maximal component of the twice dual Weyl vector. But the
twice dual Weyl vector was set by means of the procedure
 \verb"Vector" and the vector must be converted into the list to
find the maximal component of it using the standard Maple
procedure \verb "max". This action is performed by means of the
standard Maple procedure \verb"convert".
 \begin{verbatim}
  >maxel := max(convert(n, list)[]):
 \end{verbatim}

Now the matrix of the coefficients  $P_{s}^{(k)}$ can be declared.
\begin{verbatim}
 >P := array(1 .. S, 1 .. maxel):
\end{verbatim}

Let us declare the matrix of the polynomials by means of the
procedure \verb"Vector".

\begin{verbatim}
 >H := Vector[row](1 .. S):
\end{verbatim}

Each element of this matrix is defined according to the
hypothesis in the following way:
\begin{verbatim}
 >for i to S do H[i] := 1+add(P[i, k]*z^k, k = 1 .. n[i]) end do:
\end{verbatim}

It's necessary to convert the elements of the matrices $H$ and $A$
into the indexed variables for correct calculations.
\begin{verbatim}
 >for i to S do for j to S do a[i, j] := A[i, j] end do end do:
 >for i to S do h[i] := H[i] end do:
\end{verbatim}

Let us enter one more indexed variable $c_{i,v}$ for convenience.

\begin{verbatim}
 >for i to S do for v to S do c[i, v] := h[v]^(-a[i, v]) end do end do:
\end{verbatim}

We represent the set of equations as a matrix by the use of the
procedure \verb"Vector". Now the system of the differential
equations can be defined using the standard Maple procedure
\verb"diff".
\begin{verbatim}
 >equal := Vector[row](1 .. S):
 >for i to S do
 equal[i]:= diff(z*(diff(H[i],z))/H[i],z)-P[i,1]*(product(c[i,m], m = 1..S))
 end do:
\end{verbatim}

The procedure \verb"product" is the standard Maple procedure for a
product. Further we enter two more matrices for simplified
equations and  numerators of these equations using the procedure
 \verb"Vector".
 \begin{verbatim}
 > simequal := Vector[row](1 .. S):
 > newequal := Vector[row](1 .. S):
 \end{verbatim}

The first of these  matrices is filled, collecting by degrees each
of the equations by means of the standard procedures. The elements
of the second matrix are turned out by selection of the numerators
from the simplified equations.
\begin{verbatim}
 > for i to S do simequal[i] := simplify(combine(value(equal[i]), power)) end do:
 > for i to S do newequal[i] := numer(simequal[i]) end do:
\end{verbatim}

It's necessary to find out the degrees of the numerators to
collect the coefficients $P_{s}^{(k)}$ at various degrees of the
variable $z$. So let us describe a matrix, which elements are the
degrees of the variable $z$. The degrees are calculated by the
standard Maple procedure \verb"degree".
 \begin{verbatim}
 > maxcoeff := Vector[row](1 .. S):
 > for i to S do maxcoeff[i] := degree(newequal[i], z) end do:
 \end{verbatim}

We define a two-dimensional table (the standard Maple structure of
data) for the system of algebraic equations and fill the table's
elements in the following way:
\begin{verbatim}
 > coefflist := table():
 > for i to S do
    for c from 0 to maxcoeff[i] do
        coefflist[i, c] := coeff(newequal[i], z, c) = 0
    end do
  end do:
\end{verbatim}

The system should be converted into the list to solve it by means
of the standard Maple procedure \verb "solve". This action allows
us to apply the function \verb"solve".

\begin{verbatim}
 > Sys := convert(coefflist, list):
 > sol := solve(Sys):
\end{verbatim}

But the form of the answer is inconvenient. Thus
substituting the answer into the polynomial form, we get
\begin{verbatim}
 > trans := {seq(seq(P[i,j] = P[i,j], i = 1..S), j = 1..maxel)}:
 > sol := simplify(map2(subs, trans, sol)):
 > P1 := map2(subs, sol, evalm(P)):
 > for i to S do H[i]:= 1+add(P1[i,k]*z^k, k = 1..n[i]) end do;
\end{verbatim}

\begin{math}H_{1}:=1+{\displaystyle P_{1,1}z}+{\displaystyle \frac14P_{1,1}P_{2,1}z^{2}}
 +{\displaystyle
 \frac{1}{18}P_{1,1}P_{2,1}P_{3,1}z^{3}}+{\displaystyle
 \frac{1}{144}P_{1,1}P_{2,1}P_{3,1}^{2}z^{4}}+{\displaystyle
 \frac{1}{3600}P_{1,1}P_{2,1}^{2}P_{3,1}^{2}z^{5}}+{\displaystyle
 \frac{1}{129600}P_{1,1}^{2}P_{2,1}^{2}P_{3,1}^{2}z^{6}},
 \end{math}

\begin{math}
H_{2}:=1+{\displaystyle P_{2,1}z}+({\displaystyle
\frac14P_{1,1}P_{2,1}+\frac12P_{2,1}P_{3,1}})z^{2}+({\displaystyle
\frac19P_{2,1}P_{3,1}^{2}+\frac29P_{1,1}P_{2,1}P_{3,1}})z^{3}+\\
({\displaystyle \frac{1}{144}P_{2,1}^{2}P_{3,1}^{2}+\frac{1}{72}P_{1,1}P_{2,1}^{2}P_{3,1}+\frac{1}{16}P_{1,1}P_{2,1}P_{3,1}^{2}})z^{4}+
{\displaystyle
\frac{7}{600}P_{1,1}P_{2,1}^{2}P_{3,1}^{2}}z^{5}+\\
({\displaystyle
\frac{1}{1600}P_{1,1}P_{2,1}^{3}P_{3,1}^{2}+\frac{1}{5184}P_{1,1}^{2}P_{2,1}^{2}P_{3,1}^{2}+
\frac{1}{2592}P_{1,1}P_{2,1}^{2}P_{3,1}^{3}})z^{6}+({\displaystyle
\frac{1}{16200}P_{1,1}P_{2,1}^{3}P_{3,1}^{3}+\frac{1}{32400}
P_{1,1}^{2}P_{2,1}^{3}P_{3,1}^{2}})z^{7}+({\displaystyle
\frac{1}{518400}P_{1,1}P_{2,1}^{3}P_{3,1}^{4}+
\frac{1}{259200}P_{1,1}^{2}P_{2,1}^{3}P_{3,1}^{3}})z^{8}+{\displaystyle
\frac{1}{4665600}P_{1,1}^{2}P_{2,1}^{3}P_{3,1}^{4}}z^{9}+{\displaystyle
\frac{1}{466560000}P_{1,1}^{2}P_{2,1}^{4}P_{3,1}^{4}}z^{10},
 \end{math}

\begin{math}H_{3}:=1+{\displaystyle P_{3,1}z}+{\displaystyle \frac14P_{2,1}P_{3,1}z^{2}}
+({\displaystyle \frac{1}{36}P_{1,1}
P_{2,1}P_{3,1}}+{\displaystyle
\frac{1}{36}P_{2,1}P_{3,1}^{2}})z^{3}+{\displaystyle
\frac{1}{144}P_{1,1}P_{2,1}P_{3,1}^{2}z^{4}}+{\displaystyle
\frac{1}{3600}P_{1,1}P_{2,1}^{2}P_{3,1}^{2}z^{5}}+{\displaystyle
\frac{1}{129600}P_{1,1}P_{2,1}^{2}P_{3,1}^{3}z^{6}}. \end{math}

\vspace{3pt}

It should be noted that throughout the program we use a slightly
different notation for the first coefficients, i.e.
 \beq{N.0}
  P_{s,1} = P_s.
 \eeq

\section{Examples of polynomials }

 Here we present certain examples of
 polynomials corresponding to the Lie algebras
 $A_1, A_2, A_3$, $B_3$, $C_2$ and $D_4$.

 \subsection{$A_r$-polynomials, $r = 1,2,3$.}

  {\bf $A_1$-case.} The simplest example occurs in the case of
 the Lie algebra $A_1 = sl(2)$.
  We get
\cite{Iflux}
 \beq{A.1}
  H_{1}(z) = 1 + P_1 z.
 \eeq

{\bf $A_2$-case.} For the Lie algebra $A_2 = sl(3)$ with the
Cartan matrix

 \beq{A.5}
    \left(A_{ss'}\right)=
  \left( \begin{array}{*{6}{c}}
     2 & -1\\
     -1& 2\\
    \end{array}
 \right)\quad \eeq
 we have \cite{Iflux}  $n_1 = n_2 =2$ and
 \bear{A.6}
  H_{1} = 1 + P_1 z + \frac14 P_1 P_2 z^{2}, \\
  \label{A.7}
  H_{2} = 1 + P_2 z + \frac14 P_1 P_2 z^{2},
  \ear
  $s = 1,2$.

 {\bf $A_3$-case.} The polynomials for the $A_3$-case read as
 follows

 \bear{A.8}
 H_{1} = 1 + P_1 z + \frac14 P_1 P_2 z^{2} + \frac{1}{36} P_1 P_2 P_3 z^{3},\\
 \label{A.9}
 H_{2} = 1 + P_2 z + \Bigl( \frac14 P_1 P_2 + \frac14 P_2 P_3 \Bigr) z^{2} + \frac19 P_1 P_2 P_3 z^{3}
  \\ \nonumber
  + \frac{1}{144} P_1 P_2^{2} P_3 z^{4},\\
 \label{A.10}
 H_{3} = 1 + P_3 z + \frac14 P_2 P_3 z^{2} + \frac{1}{36} P_1 P_2 P_3
 z^{3}.
 \ear
  \vspace{15pt}

\subsection{$B_3$-polynomials}

 For the Lie algebra $B_3$ we get the following polynomials

\bear{B.1}
H_{1} = 1 + P_1 z + \frac14 P_1 P_2 z^{2} + \frac{1}{18} P_1 P_2 P_3 z^{3}
 + \frac{1}{144} P_1 P_2 P_3^{2} z^{4} + \frac{1}{3600} P_1 P_2^{2} P_3^{2} z^{5}
   \\ \nonumber
 + \frac{1}{129600} P_1^{2} P_2^{2} P_3^{2} z^{6},\\
 \label{B.2}
 H_{2} = 1 + P_2 z + \Bigl( \frac14 P_1 P_2 + \frac12 P_2 P_3 \Bigr) z^{2}
  + \Bigl( \frac19 P_2 P_3^{2} + \frac29 P_1 P_2 P_3 \Bigr) z^{3}
   + \Bigl( \frac{1}{144} P_2^{2} P_3^{2}
  \\
\nonumber
  + \frac{1}{72} P_1 P_2^{2} P_3 + \frac{1}{16} P_1 P_2 P_3^{2} \Bigr) z^{4}
   + \frac{7}{600} P_1 P_2^{2} P_3^{2} z^{5} +
   \Bigl( \frac{1}{1600} P_1 P_2^{3} P_3^{2} + \frac{1}{5184} P_1^{2} P_2^{2} P_3^{2}
  \\ \nonumber
   + \frac{1}{2592} P_1 P_2^{2} P_3^{3} ) z^{6} + \Bigl( \frac{1}{16200} P_1 P_2^{3} P_3^{3}
    + \frac{1}{32400} P_1^{2} P_2^{3} P_3^{2} \Bigr) z^{7} + \Bigl( \frac{1}{518400} P_1 P_2^{3} P_3^{4}
   \\ \nonumber
   + \frac{1}{259200} P_1^{2} P_2^{3} P_3^{3} \Bigr) z^{8} +
   \frac{1}{4665600} P_1^{2} P_2^{3} P_3^{4} z^{9} + \frac{1}{466560000} P_1^{2} P_2^{4} P_3^{4} z^{10},\\
   \label{B.3}
   H_{3} = 1 + P_3 z + \frac14 P_2 P_3 z^{2} + \Bigl( \frac{1}{36}P_1 P_2 P_3 + \frac{1}{36}
    P_2 P_3^{2} \Bigr) z^{3} + \frac{1}{144} P_1 P_2 P_3^{2} z^{4}
   \\ \nonumber
   + \frac{1}{3600} P_1 P_2^{2} P_3^{2} z^{5} + \frac{1}{129600} P_1 P_2^{2} P_3^{3}
   z^{6}.
\ear

\subsection{$C_2$-polynomials}

For the Lie algebra $C_2 = so(5)$ with the Cartan matrix

 \beq{C.1}
    \left(A_{ss'}\right)=
  \left( \begin{array}{*{6}{c}}
     2 & -1\\
     -2& 2\\
 \end{array}
 \right)\quad
 \eeq
 we get from  (\ref{1.4})  $n_1 = 3$ and $n_2 = 4$.

 For $C_2$-polynomials we obtain in agreement with \cite{GIM}
 \bear{C.2}
 H_1=1+P_1 z+ \frac{1}{4} P_1 P_2 z^2 + \frac{1}{36}P_1^2P_2 z^3,
 \\ \label{C.3}
  H_2=1+ P_2 z+ \frac{1}{2}P_1 P_2 z^2 +\frac{1}{9}P_1^2 P_2z^3
        +\frac{1}{144}P_1^2 P_2^2 z^4.
 \ear

 \subsection{$D_4$-polynomials}

 For the Lie algebra $D_4$ we find the following set of polynomials
 \vspace{15pt}
\bear{D.1}
H_{1} = 1 + P_1 z + \frac14 P_1 P_2 z^{2} + \Bigl( \frac{1}{36} P_1 P_2 P_3
 + \frac{1}{36} P_1 P_2 P_4 \Bigr) z^{3} + \frac{1}{144} P_1 P_2 P_3 P_4 z^{4}
\\ \nonumber
 + \frac{1}{3600} P_1 P_2^{2} P_3 P_4 z^{5} + \frac{1}{129600}P_1^{2} P_2^{2} P_3 P_4 z^{6},\\
\label{D.2}
H_{2} = 1 + P_2 z + \Bigl( \frac14 P_1 P_2 + \frac14 P_2 P_3 + \frac14 P_2 P_4 \Bigr) z^{2}
 + \Bigl( \frac19 P_1 P_2 P_3 + \frac19 P_1 P_2 P_4
 \\ \nonumber
 + \frac19 P_2 P_3 P_4 \Bigr) z^{3} + \Bigl( \frac{1}{144} P_1 P_2^{2} P_3
  + \frac{1}{144} P_1 P_2^{2} P_4 + \frac{1}{144} P_2^{2} P_3 P_4
   + \frac{1}{16} P_1 P_2 P_3 P_4 \Bigr) z^{4}
   \\ \nonumber
  + \frac{7}{600} P_1 P_2^{2} P_3 P_4 z^{5}
   + \Bigl( \frac{1}{1600} P_1 P_2^{3} P_3 P_4 +
   \frac{1}{5184} P_1 P_2^{2} P_3^{2} P_4 + \frac{1}{5184} P_1^{2} P_2^{2} P_3 P_4
    \\ \nonumber
   + \frac{1}{5184} P_1 P_2^{2} P_3 P_4^{2} \Bigr) z^{6}
    + \Bigl( \frac{1}{32400} P_1^{2} P_2^{3} P_3 P_4
    + \frac{1}{32400} P_1 P_2^{3} P_3 P_4^{2}
     + \frac{1}{32400} P_1 P_2^{3} P_3^{2} P_4 \Bigr) z^{7} \\ \nonumber
   + \Bigl( \frac{1}{518400} P_1^{2} P_2^{3} P_3 P_4^{2}
    + \frac{1}{518400} P_1^{2} P_2^{3} P_3^{2} P_4
    + \frac{1}{518400} P_1 P_2^{3} P_3^{2} P_4^{2} \Bigr) z^{8}
     \\ \nonumber
    + \frac{1}{4665600} P_1^{2} P_2^{3} P_3^{2} P_4^{2} z^{9}
     + \frac{1}{46656000} P_1^{2} P_2^{4} P_3^{2} P_4^{2} z^{10},\\
\label{D.3}
H_{3} = 1 + P_3 z + \frac14 P_2 P_3 z^{2}
+ \Bigl( \frac{1}{36} P_1 P_2 P_3 + \frac{1}{36} P_2 P_3 P_4 \Bigr) z^{3}
 + \frac{1}{144} P_1 P_2 P_3 P_4 z^{4}
  \\ \nonumber
 + \frac{1}{3600} P_1 P_2^{2} P_3 P_4 z^{5} + \frac{1}{129600} P_1 P_2^{2} P_3^{2} P_4 z^{6},\\
\label{D.4}
H_{4} = 1 + P_4 z + \frac14 P_2 P_4 z^{2} +
\Bigl( \frac{1}{36} P_1 P_2 P_4 + \frac{1}{36} P_2 P_3 P_4 \Bigr) z^{3} +
\frac{1}{144} P_1 P_2 P_3 P_4 z^{4}
  \\ \nonumber
 + \frac{1}{3600} P_1 P_2^{2} P_3 P_4 z^{5} + \frac{1}{129600} P_1 P_2^{2} P_3 P_4^{2}
 z^{6}.
\ear
 \vspace{15pt}

\section{Some relations between polynomials}

 Let us denote the set of polynomials corresponding
 to a set of parameters $P_1 > 0$, ..., $P_r > 0$ as
 following
 \beq{6.1}
   H_s = H_s(z,P_1, ...,P_r; A),
  \eeq
   $s = 1, \dots, r$,
   where $A = A[{\cal G}]$ is the Cartan matrix corresponding to
   a (semi)simple Lie algebra $\cal G$.

   \subsection{$C_{n+1}$-polynomials from $A_{2n+1}$-ones}

   The set of polynomials corresponding to the Lie algebra
   $C_{n+1}$ may be obtained from the set of polynomials
   corresponding to the Lie algebra $A_{2n+1}$
   according to the following relations
    \beq{6.AC}
   H_s(z,P_1, ...,P_{n+1}; A[C_{n+1}]) =
   H_s(z,P_1, ...,P_{n+1}, P_{n+2}= P_{n},..., P_{2n+1} = P_{1};
   A[A_{2n+1}]),
  \eeq
   $s = 1, \dots, n+1$, i.e. the parameters $P_1, ...,P_{n+1}, P_{n+2},...,
   P_{2n+1}$ are identified symmetrically w.r.t. $P_{n+1}$.
   See   Dynkin diagrams on Figs. 1-2.
   Relation (\ref{6.AC}) may be verified
   using the program from the Section 3. (For the case   $n=1$ see
   formulas from the previous section.)

   \subsection{$B_{n}$-polynomials from $D_{n+1}$-ones}

   The set polynomials corresponding to the Lie algebra
   $B_{n}$ may be obtained from the set of polynomials
   corresponding to the Lie algebra $D_{n+1}$
   according to the following relation
    \beq{6.BD}
   H_s(z,P_1, ...,P_{n}; A[B_{n}]) =
   H_s(z,P_1, ...,P_{n}, P_{n+1}= P_{n}; A[D_{n+1}]),
  \eeq
   $s = 1, \dots, n$, i.e. the parameters $P_{n}$ and $P_{n+1}$
   are identified.  See
   Dynkin diagrams on Figs. 2-3. Relation (\ref{6.BD}) may be verified
   using the program from the Section 3. (For the case   $n=3$ see
   formulas from the previous section.)

  \subsection{Reduction formulas}

  Here we denote the Cartan matrix as follows: $A = A_{\Gamma}$,
  where $\Gamma$ is the related Dynkin graph. Let $i$ be a node of
  $\Gamma$. Let us denote by $\Gamma_i$ a Dynkin graph (corresponding to a
  certain semi-simple   Lie algebra) that is obtained
  from $\Gamma$ by erasing all lines that have endpoints at $i$.
  It may be verified (e.g. by using the program) that the
  following reduction formulae are valid

  \beq{6.R1}
   H_s(z,P_1, ...,P_i=0 ,...,P_{r}; A_{\Gamma}) =
   H_s(z,P_1, ...,P_i=0 ,...,P_{r}; A_{\Gamma_i}),
  \eeq
   $s = 1, \dots, r$. Moreover,
   \beq{6.R2}
   H_i(z,P_1, ...,P_i=0 ,...,P_{r}; A_{\Gamma}) = 1.
    \eeq

   This means that  setting $P_i = 0$ we
   reduce the set of polynomials by replacing the
   the Cartan matrix $A_{\Gamma}$ by the
   Cartan matrix $A_{\Gamma_i}$. In this case the polynomial
   $H_i = 1$ corresponds to
   $A_1$-subalgebra (depicted by  the node $i$) and the parameter $P_i = 0$.
   \fnm[3]\fnt[3]{The analytical proof of the relations
   (\ref{6.AC})-(\ref{6.R1}) will be given in a separate publication}.

   As an example of reduction formulas we
   present the following relations
   \beq{6.R3}
   H_s(z,P_1, ...,P_n, P_{n+1} = 0; A[{\cal G}]) =
   H_s(z,P_1, ...,P_{n}; A[A_n]),
  \eeq
   $s = 1, \dots, n$, for ${\cal G} = A_{n+1}, B_{n+1}, C_{n+1}, D_{n+1}$
   with appropriate restrictions on $n$ (see (\ref{Ap.9})). In
   writing relation (\ref{6.R1}) we use the numbering of nodes in
   agreement with the Dynkin diagrams depicted on Figs. 1-3.

   The reduction formulas  (\ref{6.R2}) for $A_5$-polynomials
   with $P_3 = 0$ are depicted on Fig. 4. The reduced polynomials
   are coinciding with those corresponding to semisimple
   Lie algebra $A_2 \bigoplus A_1 \bigoplus A_2$.

   \begin{center}
\bigskip
\begin{picture}(66,11)
\put(5,10){\circle*{2}} \put(15,10){\circle*{2}}
\put(25,10){\circle*{2}} \put(35,10){\circle*{2}}
\put(45,10){\circle*{2}} \put(5,10){\line(1,0){10}}
\put(5,5){\makebox(0,0)[cc]{1}}
\put(15,5){\makebox(0,0)[cc]{2}}
\put(25,5){\makebox(0,0)[cc]{3}}
\put(35,5){\makebox(0,0)[cc]{$4$}}
\put(45,5){\makebox(0,0)[cc]{$5$}} \put(35,10){\line(1,0){10}}
\end{picture} \\[5pt]
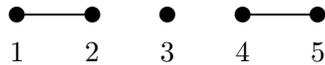
\small Fig. 4. \it  Dynkin diagram for semisimple
 Lie algebra $A_2 \bigoplus A_1 \bigoplus A_2$
  describing the set of $A_5$-polynomials with $P_3 =0$
\end{center}

 \section{\bf Conclusions}

  Here we have presented a description of computational
  program (written in Maple) for calculation of  fluxbrane polynomials
  related to classical simple Lie algebras. (Generalization to semisimple
  Lie algebras is a straightforward one.)   This program gives by product
  a verification of the conjecture suggested previously in \cite{Iflux}.
  The polynomials considered above define special solutions to open Toda chain
  equations corresponding to simple Lie algebras that may be of interest for
  certain applications of Toda chains.

  We have also considered (without proof) certain relations  between polynomials,
  e.g. so-called  reduction formulas. These relations tells us that the most important is
  the calculation of $D_n$-polynomials, since all other polynomials
  (e.g. $A_n$-,$B_n$- and $C_n$-ones) may be obtained from
  $D_n$-series of polynomials by using certain reduction formulas.

  A calculation of polynomials corresponding
  to exceptional Lie algebras (i.e. $G_2$, $F_4$, $E_6$, $E_7$ and $E_8$)
  will be considered in a separate  publications.
  (The $G_2$-polynomials were obtained earlier in \cite{GIM}.)

 \begin{center}
 {\bf Acknowledgments}
 \end{center}

 This work was supported in part by the Russian Foundation for
 Basic Research grant  Nr. $07-02-13624-ofi_{ts}$ and
 by a grant of People Friendship University (NPK MU).


 \end{document}